\documentclass[aps,prl,floatfix,twocolumn,nofootinbib]{revtex4}
\usepackage{amssymb}
\usepackage{amsmath}
\usepackage{amsfonts}
\usepackage{graphicx}
\usepackage{color}
\usepackage{xspace}
\usepackage{ulem}
\usepackage{mathtools}
\usepackage{hhline}

\newcommand{\AddrAHEP}{
AHEP Group, Instituto de F\'isica Corpuscular -- 
C.S.I.C./Universitat de Val\`encia,\\
Edificio de Institutos de Paterna, Apartado 22085,
E--46071 Val\`encia, Spain
}

\newcommand{\AddrUCL}{
Department of Physics and Astronomy, University College London,\\
London WC1E 6BT, United Kingdom
}

\newcommand{\AddrDortmund}{
Fakult\"at f\"ur Physik, Technische Universit\"at Dortmund,\\
44221 Dortmund, Germany
}

\begin{document}

\title{Falsifying High-Scale Baryogenesis with\\[1mm]Neutrinoless Double Beta Decay and Lepton Flavor Violation}

\author{Frank F. Deppisch} \email{f.deppisch@ucl.ac.uk}\affiliation{\AddrUCL}
\author{Julia Harz}        \email{j.harz@ucl.ac.uk}\affiliation{\AddrUCL}
\author{Martin Hirsch}     \email{mahirsch@ific.uv.es}\affiliation{\AddrAHEP}
\author{Wei-Chih Huang}    \email{wei-chih.huang@ucl.ac.uk}\affiliation{\AddrUCL}
\author{Heinrich P\"as}    \email{heinrich.paes@uni-dortmund.de}\affiliation{\AddrDortmund}

\begin{abstract}
Interactions that manifest themselves as lepton number violating processes at low energies in combination with sphaleron transitions typically erase any preexisting baryon asymmetry of the Universe. In this article, we discuss the constraints obtained from an observation of neutrinoless double beta decay in this context. If a new physics mechanism of neutrinoless double beta decay other than the standard light neutrino exchange is observed, typical scenarios of high-scale baryogenesis will be excluded unless the baryon asymmetry is stabilized via some new mechanism. We also sketch how this conclusion can be extended beyond the first lepton generation by incorporating lepton flavor violating processes. 
\end{abstract}

\maketitle

\section{Introduction}
The discovery of neutrino masses is typically understood as a hint for physics beyond the Standard Model~(SM). The question whether lepton number is conserved or broken is intimately related to this link. After all, neutrino masses can be realized in two different ways, either as Majorana or as Dirac masses, where in the latter case lepton number has to be protected via a newly invoked symmetry. In the following we will argue that low energy lepton number violation~(LNV) in the form of neutrinoless double beta~($0\nu\beta\beta$) decay will have far-reaching consequences for the mechanism of baryogenesis. We will also assess the impact of low energy lepton flavor vio\-lation~(LFV) to extend the argument beyond the first lepton generation.

The observed baryon asymmetry of the Universe, expressed as the baryon-to-photon number density ratio \cite{Planck:2015xua},
\begin{align}
\label{eq:etaBobs}
  \eta_B^\text{obs} = \left(6.09 \pm 0.06\right) \times 10^{-10},
\end{align}
cannot be understood within the SM~\cite{Buchmuller:2005eh}. Models of high-scale baryogenesis, with leptogenesis~\cite{Fukugita:1986hr} as the most popular realization, typically rely on the generation of an asymmetry in the $(B-L)$ number density, where $B$ and $L$ are the total baryon and lepton number, respectively. This involves the presence of $(B-L)$ and $CP$ violating interactions that occur out of thermal equilibrium. The produced $(B-L)$ asymmetry is then rapidly converted into the observed baryon asymmetry by SM $(B+L)$ violating sphaleron interactions above the electroweak~(EW) scale up to $\approx 10^{12}$~GeV~\cite{Kuzmin:1985mm}.

Among the possible $(B-L)$ violating interactions we concentrate on those with $\Delta L = 2$ and $\Delta B = 0$ which are most relevant for neutrino physics and especially for $0\nu\beta\beta$. If total lepton number is broken reasonably far above the electroweak scale such that light Majorana neutrino masses are induced, the low energy effects can be described by effective $\Delta L = 2$ operators of odd mass dimension. This additionally assumes that there are no other light particles beyond the SM at or below the EW scale. Up to dimension 11, all possible 129 operators are listed in~\cite{deGouvea:2007xp}, extending the previous work~\cite{Babu:2001ex}. We will concentrate on the following examples:
\begin{align}
	\label{eq:operators}
	\mathcal{O}_5    &= (L^i L^j) H^k H^l \epsilon_{ik} \epsilon_{jl},         \nonumber\\ 
	\mathcal{O}_7    &= (L^i d^c) (\bar{e^c} \bar{u^c}) H^j \epsilon_{ij},     \nonumber\\ 
	\mathcal{O}_9    &= (L^i L^j) (\bar{Q}_i \bar{u^c}) (\bar{Q}_j \bar{u^c}), \nonumber\\ 
	\mathcal{O}_{11} &= (L^i L^j) (Q_k d^c) (Q_l d^c) H_m \bar{H_i} 
	\epsilon_{jk}\epsilon_{lm}, 
\end{align}
written in terms of the SM fields $L = (\nu_L, e_L)^T$, $Q = (u_L, d_L)^T$, $H = (H^+, H^0)^T$, $e^c$, $u^c$ and $d^c$, where the fermions are described as left-handed two-component fields. The bracketing denotes the chosen Lorentz contraction and we suppress the possible flavor and color structures of the operators. The case $\mathcal{O}_{5}$ corresponds to the well-known Weinberg operator, but all the other operators will generate light Majorana neutrino masses at various loop levels after EW symmetry breaking~\cite{deGouvea:2007xp}. The above operators act as representative examples that mediate $0\nu\beta\beta$ decay via standard or nonstandard light neutrino exchange, or via short-range interactions at tree level.

\section{Neutrinoless Double Beta Decay}
\label{sec:0vbb}

\begin{figure}[t]
\centering
\includegraphics[clip,width=0.46\linewidth]{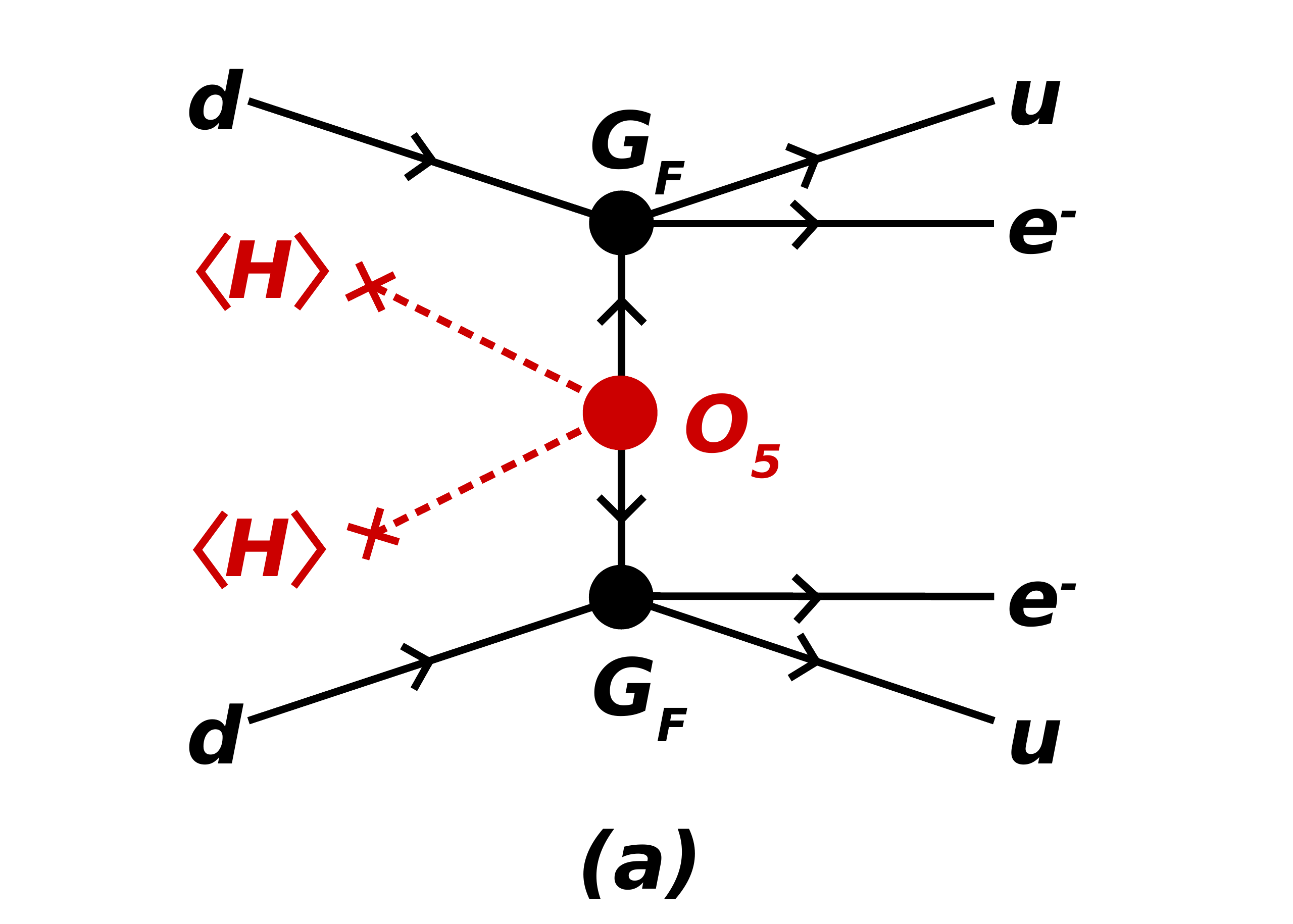}
\includegraphics[clip,width=0.46\linewidth]{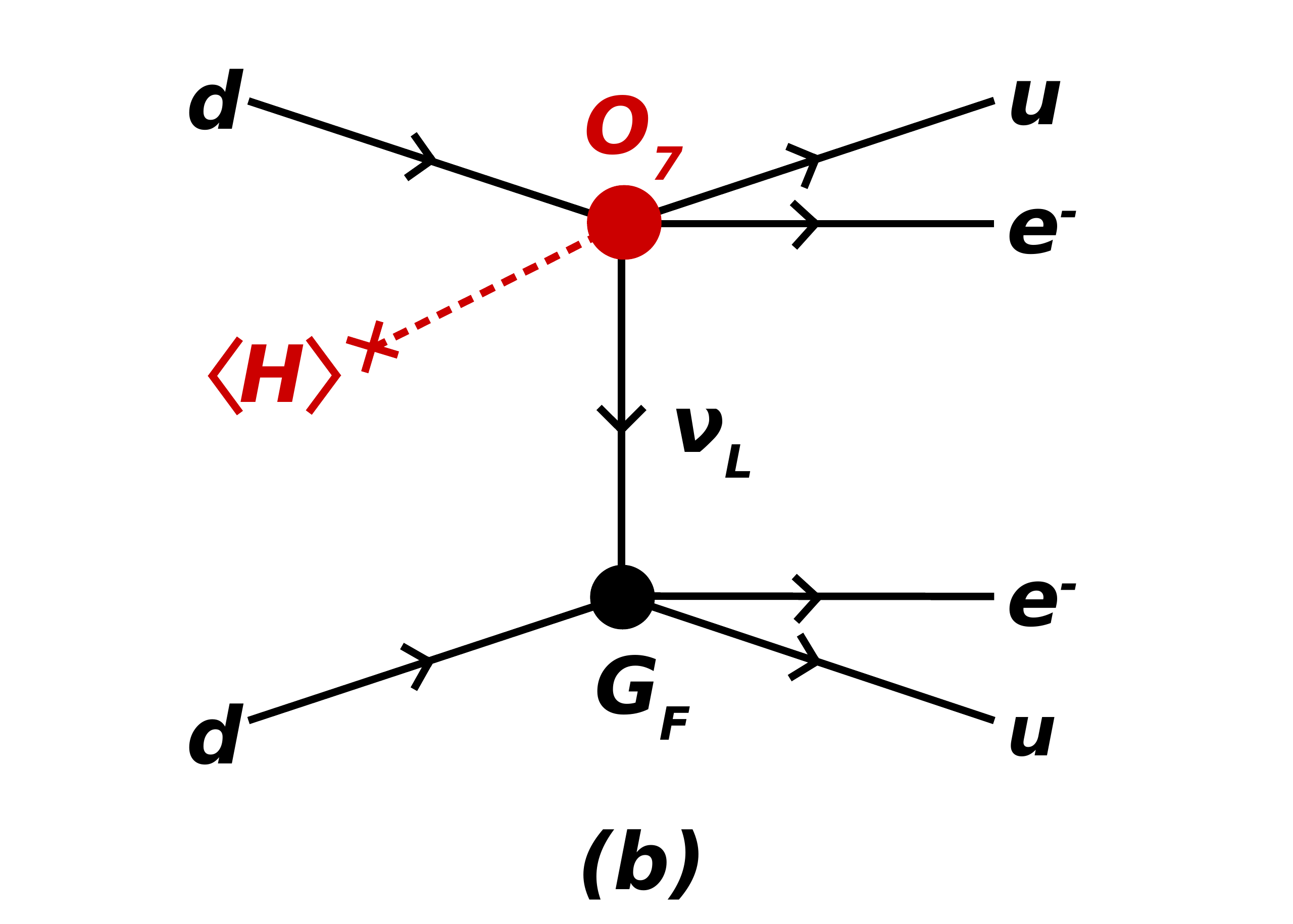}\
\includegraphics[clip,width=0.46\linewidth]{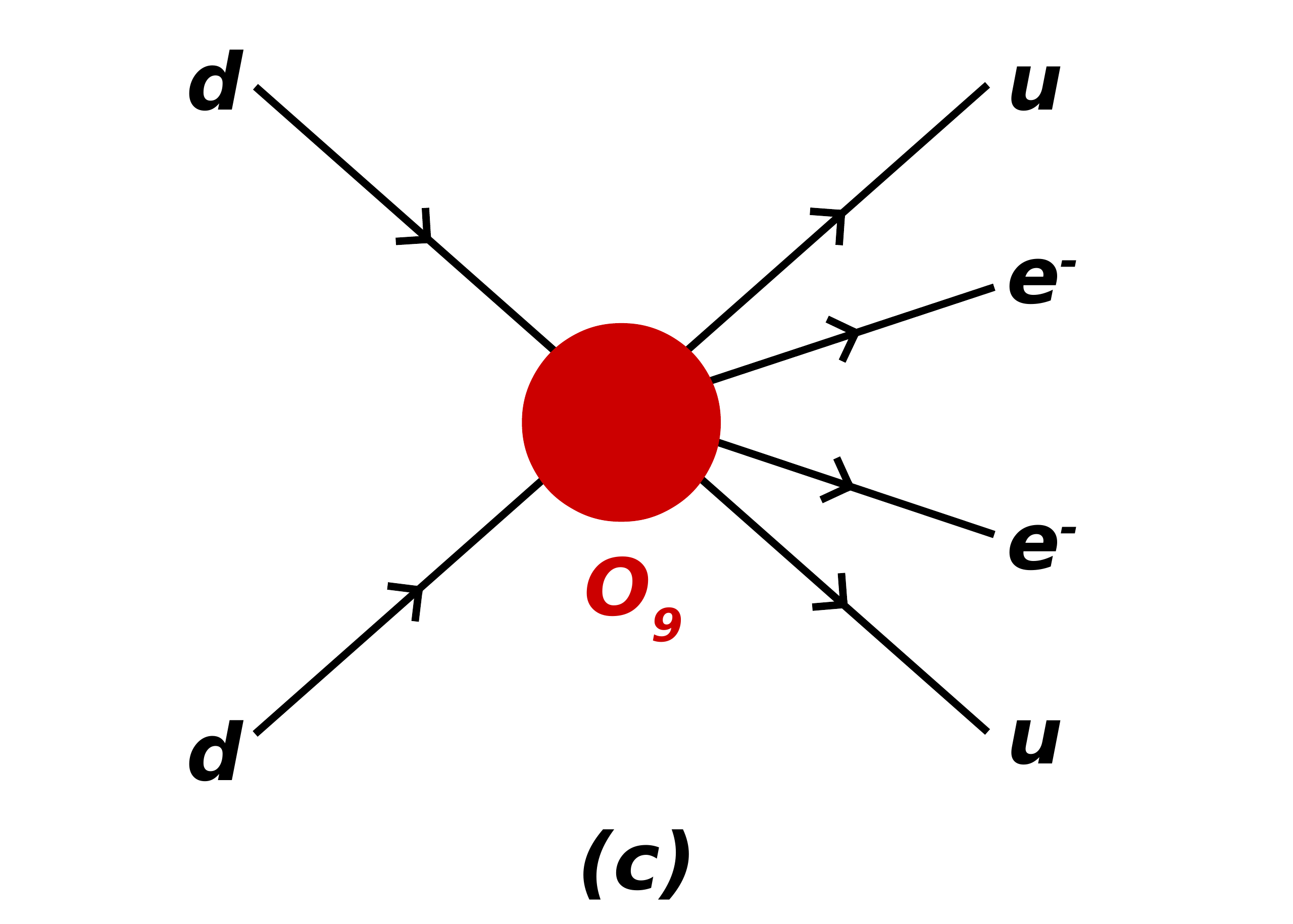}
\includegraphics[clip,width=0.46\linewidth]{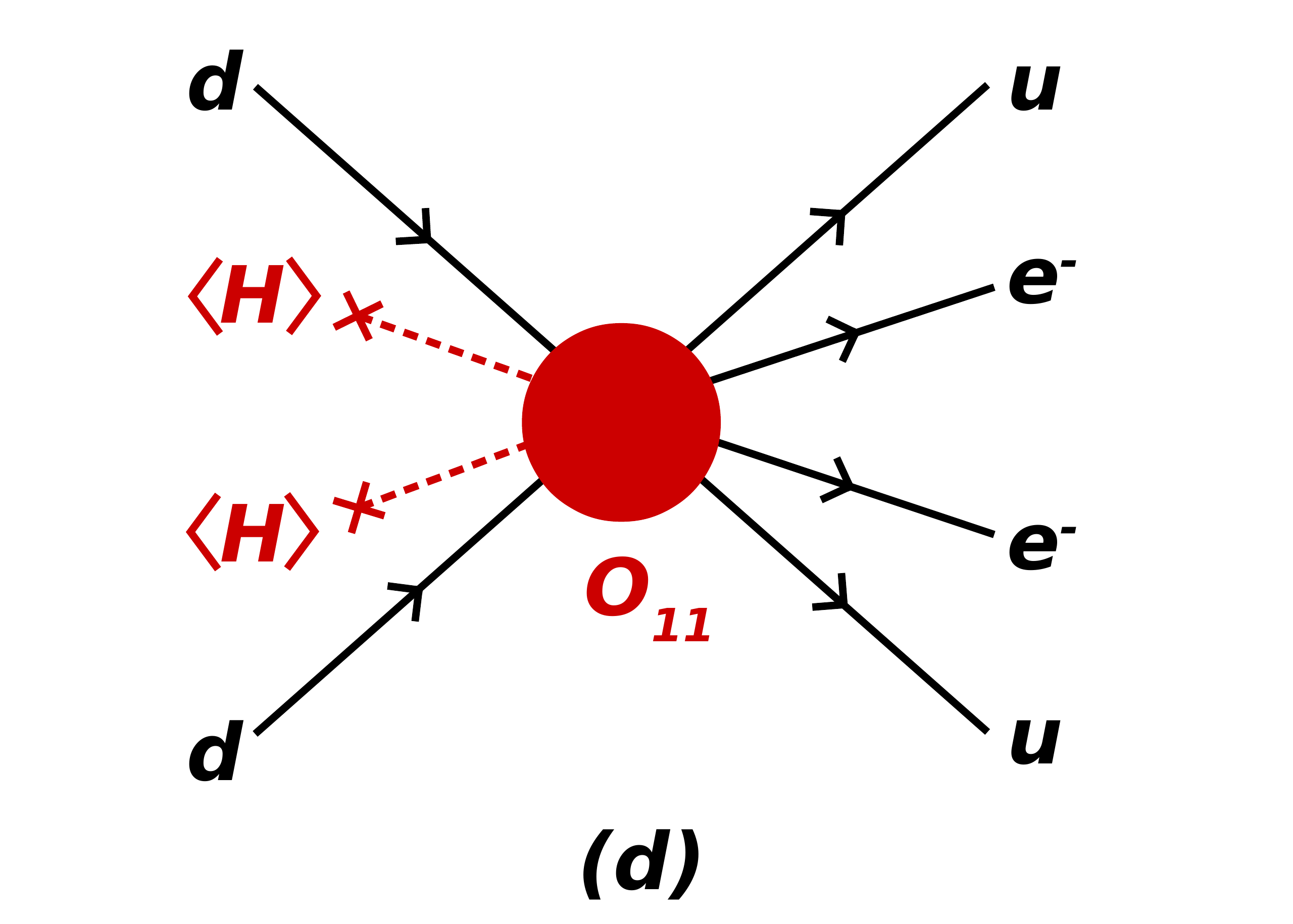}
\caption{Contributions to $0\nu\beta\beta$ decay generated by the operators $\mathcal{O}_5$ (a), $\mathcal{O}_7$ (b), $\mathcal{O}_9$ (c) and $\mathcal{O}_{11}$ (d), as given in Eq.~\eqref{eq:operators}, in terms of effective vertices, pointlike at the nuclear Fermi momentum scale.}
\label{fig:0vbbcontrib}  
\end{figure}

The most prominent probe of low energy LNV is $0\nu\beta\beta$ decay, the simultaneous transition of two neutrons into two protons and two electrons. The most general Lagrangian triggering the decay can be parametrized as depicted in Fig.~\ref{fig:0vbbcontrib}, in terms of effective 6-dim and 9-dim operators at the nuclear Fermi scale ${\cal O}(100\text{ MeV})$~\cite{Pas:1999fc}. The diagrams show the exchange of a light Majorana neutrino generated by $\mathcal{O}_5$ between two SM Fermi interactions~(a), the exchange of a light neutrino between a Fermi interaction and the operator $\mathcal{O}_7$~(b), and two short-range contributions triggered by the operators $\mathcal{O}_9$~(c) and $\mathcal{O}_{11}$~(d).

The $0\nu\beta\beta$ half-life can be succinctly written in terms of an effective coupling $\epsilon_i$ of a single operator as $T_{1/2}^{-1} = \epsilon_i^2 G_i |M_i|^2$, where $G_i$ and $M_i$ are the nuclear $0\nu\beta\beta$ phase space factor and matrix element, respectively, for a given isotope and operator. The effective couplings $\epsilon_i$ are connected to the scales of the operators in Eq.~\eqref{eq:operators} as~\cite{Deppisch:2012nb}
\begin{gather}
\label{eq:epsilons}
	                m_e \epsilon_5          = \frac{g^2 v^2}{\Lambda_5},   \,\,
	\frac{G_F \epsilon_7}{\sqrt{2}}         = \frac{g^3 v}{2 \Lambda_7^3}, \,\,
	\frac{G_F^2 \epsilon_{\{9,11\}}}{2 m_p} = \{\frac{g^4}{\Lambda_9^5},
	                                            \frac{g^6 v^2}{\Lambda_{11}^7} \}.
\end{gather}
In terms of the effective $0\nu\beta\beta$ mass $m_{ee}$, one simply has $\epsilon_5 = m_{ee} / m_e$ with the electron mass $m_e$, whereas the other couplings are normalized with respect to the Fermi coupling $G_F$ and the proton mass $m_p$. The Higgs vacuum expectation value $v = 174$~GeV arises from EW symmetry breaking thereby generating the effective 6-dim and 9-dim operators for $0\nu\beta\beta$. Powers of a generic (average) coupling constant $g$ are included to illustrate the scaling expected in a tree level ultraviolet (UV) completion of an operator. In the following we will set $g = 1$ for simplicity.

The most stringent bounds are currently derived from experimental $0\nu\beta\beta$ searches in $^{76}$Ge and $^{136}$Xe with 90\% C.~L. limits of $T_{1/2} > 2.1 \times 10^{25}$~y~\cite{Agostini:2013mzu} and $T^0_{1/2} > (1.1 - 1.9) \times 10^{25}$~y~\cite{Albert:2014awa, Gando:2012zm}, respectively. In deriving the corresponding scales of the operators we use the results of \cite{Deppisch:2012nb} for $^{76}$Ge. Planned future experiments aim to increase the sensitivity by potentially two orders of magnitude to $T_{1/2} \approx 10^{27}$~y \cite{Gomez-Cadenas:2015twa}. Assuming the dominance of a single operator, the half-life can be expressed as
\begin{align}
	\label{eq:linkdim}
	T_{1/2} = 2.1 \times 10^{25} \text{ y} \cdot
	\left(\Lambda_D / \Lambda_D^0 \right)^{2d-8},
\end{align}
where $\Lambda_D^0$ is the scale corresponding to the current sensitivity. Table~\ref{tab:LNVoverview} lists the values of $\Lambda_D^0$ for our selection of operators. The scaling dimension $d$ is identical to the operator dimension $D$ if $0\nu\beta\beta$ is generated at tree level from the underlying operator, as in the cases we discuss, but could be smaller for loop-induced diagrams. As mentioned before, the operators in Eq.~\eqref{eq:operators} act as examples for the different types of $0\nu\beta\beta$ decay mediation. Similar results hold for the other 125 operators and other Lorentz structures. The latter will affect the $0\nu\beta\beta$ sensitivity somewhat, but due to the high dimensionality of the operators this will only weakly impact the derived scales. Many of the 129 operators will induce $0\nu\beta\beta$ nonstandard mechanisms only at the loop level; in such cases, there will be additional loop suppression factors in the relations analogous to Eq.~\eqref{eq:epsilons}. This will make it unlikely that such contributions can be observed in $0\nu\beta\beta$ decay, but if they were, our following argumentation with respect to baryogenesis would be even stronger.

If $0\nu\beta\beta$ decay was observed, the responsible operator would still be unknown. Although discriminating between the different underlying operators is a challenging task, various ideas have been proposed regarding how this could be achieved, at least for a subset of the various contributions.
Cosmological observations such as anisotropies of the cosmic microwave background or the large scale structure, can set stringent constraints on the sum of neutrino masses; the Planck Collaboration, for example, recently attained $\sum m_\nu < 0.17~\mathrm{eV}$~\cite{Planck:2015xua}, which can be further improved by future experiments~\cite{Lesgourgues:2012uu}. An inconsistency between the neutrino masses determined by cosmology and an observed $0\nu\beta\beta$ decay, would rule out the standard interpretation with three light neutrinos and could therefore point us to a nonstandard contribution to $0\nu\beta\beta$ decay.

Different mechanisms can be directly distinguished in searches for $0\nu\beta\beta$ decay by looking at the kinematic distribution of the outgoing electrons~\cite{Doi:1982dn}. This technique will be used in the future SuperNEMO experiment~\cite{Pahlka:2008dw}. As it will be able to measure the angular and energy distribution of the electrons, it can identify long-range contributions, such as the $\mathcal{O}_7$ discussed above, leading to right-handed currents. Whereas for the standard mass mechanism with $V-A$ couplings the two electrons are expected to be preferably emitted back to back with comparable energies, the contribution of $\mathcal{O}_7$ will lead to a signal with the electrons being preferably emitted in the same direction with one taking most of the energy~\cite{Doi:1982dn}. SuperNEMO has an expected sensitivity of $T_{1/2} = 1.2\times 10^{26}$~y in the case of the standard light neutrino exchange and $T_{1/2} = 6.1\times 10^{25}$~y if $0\nu\beta\beta$ decay is mediated by a right-handed current generated from the operator $\mathcal{O}_7$, for a 500~kg~y exposure to the isotope $^{82}$Se~\cite{Arnold:2010tu}. If $0\nu\beta\beta$ decay is discovered in any of the ongoing or planned experiments, it will be important to improve such experimental techniques further as this method can directly test for the presence of nonstandard right-handed current contributions.  

Another possibility is to compare $0\nu\beta\beta$ decay rates of different isotopes as shown in Ref.~\cite{Deppisch:2006hb}. As in the half-life ratio of two different isotopes the new physics parameters drop out; it depends only on the nuclear matrix elements and phase space integrals which in turn depend on the underlying mechanism. When comparing the experimentally determined ratio with the theoretical prediction, potential new physics contributions can be determined. Further possibilities include the comparison of $0\nu\beta^- \beta^-$ decay with $0\nu\beta^{+}/\mathrm{EC}$~\cite{Hirsch:1994es}, the comparison of $0\nu\beta^-\beta^-$ and $0\nu\beta^+\beta^+$ decay~\cite{Hirsch:1994es} as well as the comparison between $0\nu\beta\beta$ decay to the ground state and an excited state~\cite{Simkovic:2001qf}.

Moreover, as has been discussed in Ref.~\cite{Helo:2013dla}, short-range contributions that lead to the LHC analog of neutrinoless double beta decay feature characteristic observables. These include invariant mass peaks corresponding to the masses of heavy particles created on-shell, and a possible asymmetry in the rate of the processes with $e^+e^+$ and $e^-e^-$  final states which could allow for a very concrete identification of the new physics scenario.

Thus, various ideas exist to achieve the necessary discrimination between the standard mechanism from other contributions, which is crucial for our analysis due to the large disparity of scales, cf. Table~\ref{tab:LNVoverview}.

\section{Lepton Flavor Violation}
\label{sec:lfv}

Neutrinoless double beta decay can only probe the electron-electron component of the LNV operators discussed above, which should in general be dressed with appropriate coefficients in flavor space; for example, $1 / \Lambda_9^5 \to c_{\alpha\beta} / \Lambda_9^5$ (suppressing quark flavor) with $\alpha,\beta = e,\mu,\tau$ and $c_{\beta\alpha} = c_{\alpha\beta}$. Observation of only $0\nu\beta\beta$ does not allow us to model independently fix any of the coefficients except $c_{ee}$. LNV meson decays and direct searches at the LHC might probe the corresponding $\mu\mu$ and $\tau\tau$ coefficients, whereas the off-diagonal transitions violate not only the total lepton number but also individual flavor numbers by one unit. While LFV and LNV can be observed simultaneously in certain processes, such as $\mu^+ \to e^-$ conversion in nuclei or in direct searches at the LHC, the most stringent limits on LFV are set on 6-dim $\Delta L = 0$ operators of the form $\mathcal{O}_{\ell\ell\gamma} = \mathcal{C}_{\ell\ell\gamma} \bar L_\ell \sigma^{\mu\nu} \bar{\ell^c} H F_{\mu\nu}$ and $\mathcal{O}_{\ell\ell q q} = \mathcal{C}_{\ell\ell q q} (\bar{\ell} \, \Pi_1 \ell) (\bar{q} \, \Pi_2 q)$ (the $\Pi_i$ represent possible Lorentz structures)~\cite{Raidal:2008jk}, with $\ell=e,\mu,\tau$. We define the LFV operator scales $\Lambda_i$ as
\begin{gather}
	\mathcal{C}_{\ell\ell\gamma} = \frac{e g^3}{16\pi^2 \Lambda^2_{\ell\ell\gamma}}, \quad
	\mathcal{C}_{\ell\ell q q}    = \frac{g^2}{\Lambda^2_{\ell\ell q q}},
\end{gather}
again keeping track of generic couplings in a UV complete model through powers of $g$, which we set to unity in our numerical results. The operator $\mathcal{O}_{\ell\ell\gamma}$ necessarily involves an electromagnetic coupling and cannot be induced at tree level. We therefore include the elementary charge $e$ and a loop suppression factor in $\mathcal{C}_{\ell\ell\gamma}$. We do not assume any correlation between the 6-dim $\Delta L = 0$ operators and the $\Delta L = 2$ operators discussed above. Instead, each operator can live at a different scale, only constrained or fixed by the experimental data. This will allow us to infer for what temperatures the individual lepton flavor asymmetries are in equilibrium.  

Among the possible low energy LFV processes, we consider the following observables along with their current limits at 90\% C.~L.: the decay branching ratios $\text{Br}_{\mu\to e \gamma} < 5.7\times 10^{-13}$~\cite{Adam:2013mnn}, $\text{Br}_{\tau\to \ell\gamma} \lesssim 4.0\times 10^{-8}$ ($\ell=e,\mu$)~\cite{Agashe:2014kda} and the $\mu-e$ conversion rate $\text{R}^\text{Au}_{\mu\to e} < 7.0\times 10^{-13}$~\cite{Agashe:2014kda}. The expected sensitivities of ongoing and planned experiments are $\text{Br}_{\mu\to e \gamma} \approx 6.0\times 10^{-14}$~\cite{Baldini:2013ke}, $\text{Br}_{\tau\to \ell\gamma} \approx 1.0\times 10^{-9}$~\cite{Aushev:2010bq} and $\text{R}^\text{Al}_{\mu\to e} \approx 2.7\times 10^{-17}$~\cite{TDRComet}. Similar to Eq.~\eqref{eq:linkdim} we relate the observables with the corresponding operator scales, for example for $\mathcal{O}_{\mu e\gamma}$,
\begin{align}
\label{eq:linkdimLFV}
	\text{Br}_{\mu\to e\gamma} = 5.7\times 10^{-13} \cdot 
	\left( \Lambda_{\mu e\gamma}^0 / \Lambda_{\mu e\gamma} \right)^4,
\end{align}
and analogously for the operators $\mathcal{O}_{\tau\ell\gamma}$ and $\mathcal{O}_{\mu eqq}$. 
We omit the possibility that the LFV processes are induced by higher dimensional operators. 
The scales $\Lambda_i^0$ corresponding to the current sensitivities are shown in Table~\ref{tab:LFVoverview}.
\begin{table}[t!]
\begin{tabular}{lll}
\hline
$\mathcal{O}_D$    & ~~$\lambda^0_D$ [GeV]  & ~~$\Lambda^0_D$ [GeV]\\
\hline
$\mathcal{O}_5$    & ~~$9.2 \times 10^{10}$ & ~~$9.1\times 10^{13}$ \\
$\mathcal{O}_7$    & ~~$1.2 \times 10^{2}$  & ~~$2.6\times 10^{4}$  \\
$\mathcal{O}_9$    & ~~$4.3 \times 10^{1}$  & ~~$2.1\times 10^{3}$  \\
$\mathcal{O}_{11}$ & ~~$7.8 \times 10^{1}$  & ~~$1.0\times 10^{3}$  \\
\hline
\end{tabular}
\caption{Operator scale $\Lambda^0_D$ and minimal washout scale $\lambda_D^0$ for the LNV operators in Eq.~\eqref{eq:operators} and the current $0\nu\beta\beta$ sensitivity $T_{1/2} = 2.1\times 10^{25}$~y.}
\label{tab:LNVoverview}
\end{table}
%

\section{Lepton Asymmetry Washout}
\label{sec:washout}

We now consider the washout of a preexisting net lepton asymmetry from the above operators. Including only the washout processes generated by a single $D$-dimensional LNV operator, the Boltzmann equation for the net lepton number $\eta_L$, normalized to the photon density $n_\gamma$, can be expressed as
\begin{align}
\label{eq:be}
	 n_\gamma H T \frac{d\eta_L}{d T} = c_D \frac{T^{2D-4}}{\Lambda_D^{2D-8}} \eta_L.
\end{align}
Here, the equilibrium photon density is $n_\gamma \approx 2T^3/\pi^2$, the Hubble parameter is $H \approx 1.66 \sqrt{g_*} T^2 / \Lambda_\text{Pl}$, with the effective number of relativistic degrees of freedom $g_*$ ($\approx 107$ in the SM) and the Planck scale $\Lambda_\text{Pl} = 1.2 \times 10^{19}$~GeV. The constant $c_D$ is calculated for each operator by determining the scattering density integrated over the whole phase space and summing over all possible initial and final states. For the operators in Eq.~\eqref{eq:operators} it is given by $c_{\{5,7,9,11\}} = \{ 8/\pi^5, 27/(2\pi^7), 3.2\times 10^4/\pi^9, 3.9\times 10^5/\pi^{13}\}$.

The $\Delta L = 2$ processes induced by the operator $\mathcal{O}_D$ can be considered to be in equilibrium and washout of the lepton asymmetry is effective if 
\begin{align}
\label{eq:washout}
  \frac{\Gamma_W}{H} &\equiv \frac{c_D}{n_\gamma H}\frac{T^{2D-4}}{\Lambda_D^{2D-8}} 
	= c_D' \frac{\Lambda_\text{Pl}}{\Lambda_D}\left(\frac{T}{\Lambda_D}\right)^{2D-9} \gtrsim 1,
\end{align}
with $c_D' = \pi^2 c_D/(3.3 \sqrt{g_*}) \approx 0.3 c_D$. This is the case in the temperature interval
\begin{align}
\label{eq:temp_limit}
  \Lambda_D \left( \frac{\Lambda_D}{c_D' \Lambda_\text{Pl}} \right)^{\frac{1}{2D-9}} 
	\equiv \lambda_D \lesssim T \lesssim \Lambda_D.
\end{align}
The upper limit $T \lesssim \Lambda_D$ is to ensure that only scales are considered where the effective operator approach is valid. Around the temperature $T \approx \Lambda_D$ it will become necessary to consider the underlying model with the general effect that the washout rate will be regularized by the exchange of heavy particles with a mass scale of $\Lambda_D$.

\begin{figure}[t!]
\centering
\includegraphics[clip,width=0.9\linewidth]{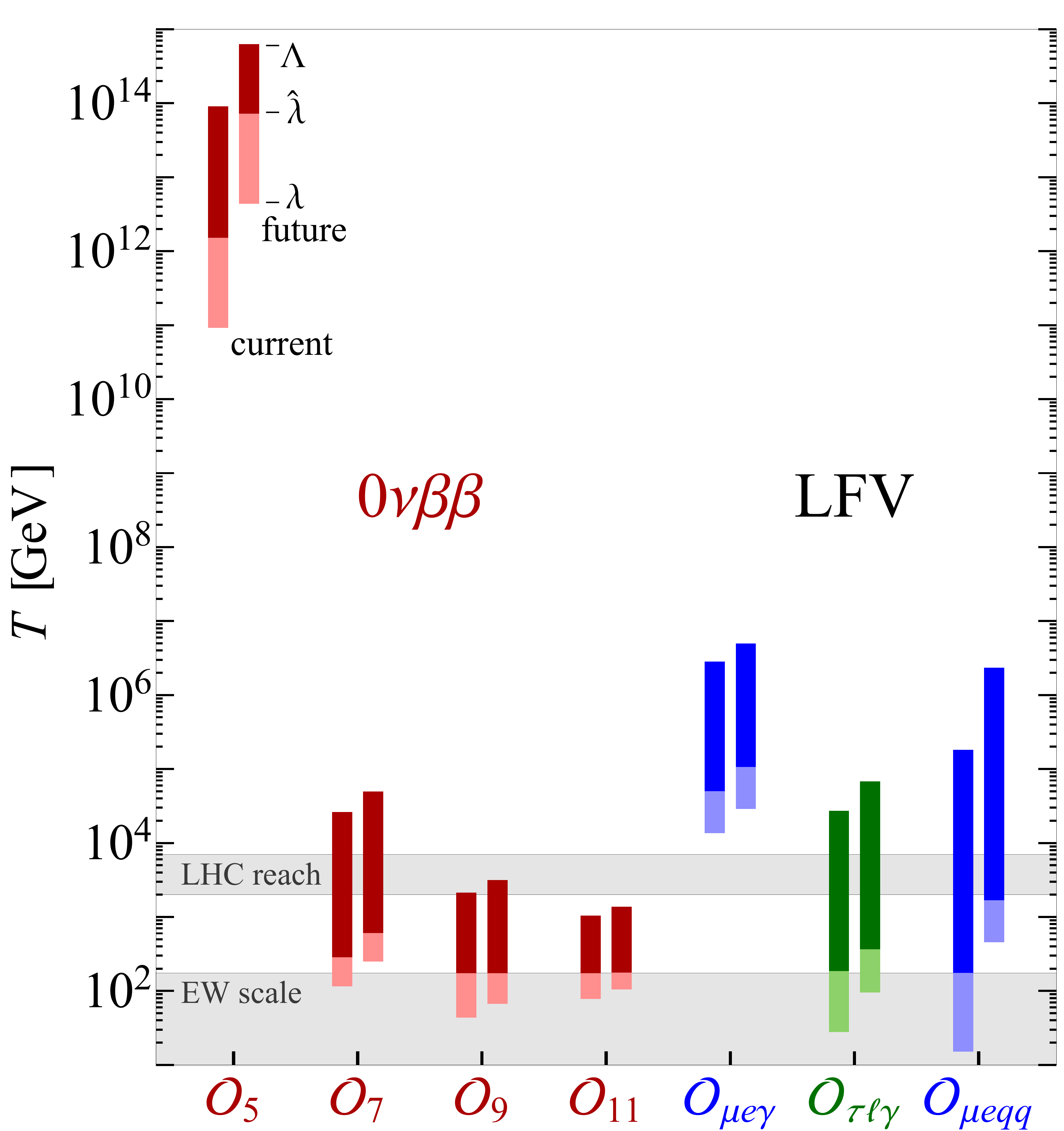}
\caption{Temperature intervals where the given LNV and LFV operators are in equilibrium, defined by the operator scale $\Lambda$ and the minimal washout scales $\lambda, \hat\lambda$ as described in the text. In each case, for the left (right) bar it is assumed that the corresponding process is observed at the current (future) experimental sensitivity as given in the text.}
\label{fig:ranges}  
\end{figure}

\begin{table}[t!]
\begin{tabular}{lll}
\hline
$\mathcal{O}_i$                & ~~$\lambda^0_i$ [GeV] & ~~$\Lambda^0_i$ [GeV] \\
\hline
$\mathcal{O}_{\mu e\gamma}$    & ~~$1.4 \times 10^{4}$ & ~~$2.8\times 10^{6}$  \\
$\mathcal{O}_{\tau\ell\gamma}$ & ~~$2.8 \times 10^{1}$ & ~~$2.7\times 10^{4}$  \\
$\mathcal{O}_{\mu eqq}$        & ~~$1.5 \times 10^{1}$ & ~~$1.8\times 10^{5}$  \\
\hline
\end{tabular}
\caption{As Table~\ref{tab:LNVoverview} but for the given LFV operators, using the current sensitivity of their respective observable.}
\label{tab:LFVoverview}
\end{table}
An asymmetry generated at scales above $\lambda_D$ will be washed out if $0\nu\beta\beta$ is observed at a corresponding rate and if it could be established that the operator in question gives the dominant contribution. Table~\ref{tab:LNVoverview} shows the values of $\lambda^0_D$ for the operators in Eq.~\eqref{eq:operators} and the current experimental $0\nu\beta\beta$ sensitivity. The determination of the lower limit on the scale of baryogenesis can be made more precisely by solving the Boltzmann equation \eqref{eq:be} to determine the suppression of a primordial asymmetry down to the EW scale where any remainder is converted to a baryon asymmetry by sphaleron processes. This leads to the increased lower limit
\begin{align}
	\hat\lambda_D \approx 
	\left[(2D-9) \ln\left(\frac{10^{-2}}{\eta_B^\text{obs}}\right) \lambda_D^{2D-9} + v^{2D-9}\right]^{\frac{1}{2D-9}}\!\!\!,
\end{align}
where we conservatively assume a primordial asymmetry of order one, perhaps generated in a non-thermal fashion. The effective washout intervals for the different operators are shown in Fig.~\ref{fig:ranges}, for both the current and the future experimental $0\nu\beta\beta$ sensitivity $T_{1/2} = 10^{27}$~y.

An analogous analysis can be applied to the $\Delta L = 0$ LFV operators, but instead of leading to a washout of a net lepton number, we are here interested in the temperature interval where two individual flavor number asymmetries are equilibrated by LFV processes. When this interval overlaps with the $\Delta L = 2$ washout interval of one net flavor number (i.e. electron number if $0\nu\beta\beta$ is observed), the net number of the other flavor will be efficiently washed out as well. Table~\ref{tab:LFVoverview} shows the corresponding lower limits for effective washout based on the current experimental sensitivities. Furthermore, the LFV equilibrium intervals are displayed in Fig.~\ref{fig:ranges} in comparison with the LNV washout intervals, for both the current and expected future sensitivities. 

The most immediate feature in Table~\ref{tab:LNVoverview} and Fig.~\ref{fig:ranges} is the stark dichotomy between the scale of the operator $\mathcal{O}_5 \approx 10^{14}$~GeV and the scales of the other LNV operators $\approx 10^{3-5}$~GeV. In this way, $0\nu\beta\beta$ decay probes both very high scales and the TeV scale. Our main conclusion is that if $0\nu\beta\beta$ decay is observed and triggered by an operator other than $\mathcal{O}_5$, the resulting washout would rule out baryogenesis mechanisms above the corresponding scale $\hat\lambda_D$ and therefore essentially anywhere but close to the EW scale. We want to stress that the strong washout intervals only apply if $0\nu\beta\beta$ is actually observed; if no signal is seen one may only conclude that the washout of the corresponding operator is weak below $\hat\lambda_D$. Future experimental improvements of the $0\nu\beta\beta$ sensitivity may not be able increase much the reach in the scales of $\mathcal{O}_{7,9,11}$ but they can still extend to probe the phenomenologically interesting Terascale up to $\approx 50$~TeV. 

Future LFV searches will probe intermediate scales up to $\approx 10^6$~GeV, but if $\tau\to \ell\gamma$ or $\mu-e$ conversion in nuclei was observed, the involved flavors would be equilibrated around the same temperatures as the washout from the LNV operators $\mathcal{O}_{7,9,11}$. Combining $0\nu\beta\beta$ and LFV searches can therefore have a potentially strong impact on our understanding of the baryon asymmetry of the Universe. On the other hand, the limit on $\text{Br}_{\mu\to e\gamma}$ is already so severe that there is essentially no overlap with the LNV ranges. We also show $\hat\lambda$ for the LFV operators but it should not be interpreted as a lower limit on effective scattering; it merely indicates that the flavor equilibration weakens as the temperature decreases.

\section{Conclusions}
\label{sec:conclusions}

Our results demonstrate that an observation of $0\nu\beta\beta$ decay can impose a stringent constraint on mechanisms of high-scale baryogenesis. More concretely, if $0\nu\beta\beta$ decay is triggered by any nonstandard mechanism, Fig.~\ref{fig:0vbbcontrib}~(b), \ref{fig:0vbbcontrib}~(c), and \ref{fig:0vbbcontrib}~(d), high-scale baryogenesis is generally excluded. For contribution~(b) an experiment such as SuperNEMO which is sensitive to electron tracks and momenta may be able to discriminate this case from the mass mechanism. For contributions (c) and (d), an observation of $0\nu\beta\beta$ decay will typically also imply the observation of LNV processes at the LHC~\cite{Helo:2013dla, Bonnet:2012kh}. A discovery at the LHC in itself is sufficient to exclude high-scale baryogenesis scenarios~\cite{Deppisch:2013jxa}. In Fig.~\ref{fig:ranges} we indicate the approximate reach of the LHC in direct LNV searches~\cite{Deppisch:2013jxa}, illustrating that the LHC and $0\nu\beta\beta$ probe very similar scales. Given the expected high sensitivity of upcoming cosmological observations to the sum of the light neutrino masses~\cite{Lesgourgues:2012uu} at the level of the oscillation mass splitting, the comparison with $0\nu\beta\beta$ decay searches can become potentially important as well; an incompatibility within the standard 3-flavor neutrino framework could indicate the presence of a nonstandard contribution from any of the 7-, 9- or 11-dimensional operators. Since our arguments demonstrate the importance of distinguishing between mechanisms, we hope that our work motivates experimentalists to refine experimental strategies for this purpose.

However, we would like to emphasize that for our main conclusion, i.e. the falsification of baryogenesis mechanism above $\approx 500$~GeV, it is not necessary to exactly pinpoint the dominant nonstandard $0\nu\beta\beta$ operator but only to establish the presence of any nonstandard contribution in near future $0\nu\beta\beta$ decay searches. In any case, the results of this work provide further motivation to vigorously search for LNV; if observed, both the mechanism of neutrino mass generation and the then necessarily low-scale mechanism of baryogenesis could be discovered. 

Loopholes to this argument exist, such as the LNV washout not affecting a specific lepton flavor. This would be especially problematic in case of the third generation which is difficult to probe at both low and high energies. We have demonstrated that simultaneous observation of $0\nu\beta\beta$ and LFV processes can be combined to understand if individual flavor asymmetries are washed out. This represents a rather non-trivial motivation for LFV searches. Due to the presence of $(B+L)$ violating sphaleron processes, our arguments do apply to general baryogenesis mechanisms with $\Delta(B-L) \neq 0$ and not only to the case $\Delta L = 2$, $\Delta B = 0$, but models with new conserved quantum numbers or hidden sectors may be exempt~\cite{Weinberg:1980bf}. Such protection mechanisms should be addressed explicitly in any model combining low-scale LNV with high-scale baryogenesis. Apart from this caveat, our analysis is based on an effective operator approach, with the only fundamental assumption that lepton number is broken above the EW scale. It is therefore model independent and conservative, and similar bounds can be made more stringent in specific models. For example, successful baryogenesis via leptogenesis can provide bounds on the light neutrino masses in seesaw scenarios~\cite{Nelson:1990ir}.

In summary, if $0\nu\beta\beta$ decay is observed and it can be demonstrated that it is triggered by a nonstandard mechanism, baryogenesis is likely to occur at a low scale. If on the other hand, baryogenesis is a high-scale phenomenon, the only manifestation of LNV at low scales is $0\nu\beta\beta$ decay through the standard mass mechanism. In this case it is highly probable that the origin of neutrino mass generation occurs at a high scale as well.

\section{Acknowledgments}
The authors would like to thank Lukas Graf for useful discussions and for the assistance in preparing this work. The work of FFD, JH and WCH was supported partly by the London Centre for Terauniverse Studies (LCTS), using funding from the European Research Council via the Advanced Investigator Grant 267352. MH is supported by the Spanish grants FPA2014-58183-P and Multidark CSD2009-00064 (MINECO), and PROMETEOII/2014/084 (Generalitat Valenciana).

\end{document}